\documentclass[11pt]{article}
\usepackage[final]{acl}
\usepackage{times}
\usepackage{latexsym}
\usepackage[T1]{fontenc}
\usepackage[utf8]{inputenc}
\usepackage{microtype}
\usepackage{inconsolata}
\usepackage{graphicx}
\usepackage{float}
\usepackage{stfloats}
\usepackage{makecell}
\usepackage{amsmath}
\usepackage{amssymb}
\usepackage{booktabs}
\usepackage{multirow}
\usepackage{tabularx}
\usepackage{xcolor}
\usepackage{subcaption}
\usepackage{tipa}
\makeatletter
\g@addto@macro{\UrlBreaks}{\UrlOrds}
\makeatother

\hypersetup{
  pdftitle={SpeechSplice: Pronunciation Control in Encoder-Decoder LLM-based TTS by Replacing Encoder Hidden States},
  pdfauthor={Kiyoshi Kurihara, Jumpei Ukita, Yusuke Iwasawa, Yutaka Matsuo}
}

\newcommand{\urlfootnote}[2]{\footnote{\begingroup\raggedright\href{#2}{#1}\par\endgroup}}
\newcommand{\releaseurl}{\url{https://github.com/Kuri-san-jp/speechsplice}}

\title{SpeechSplice: Pronunciation Control in Encoder--Decoder LLM-based TTS \\ by Replacing Encoder Hidden States
}

\author{
  \textbf{Kiyoshi Kurihara\textsuperscript{1,2}},
  \textbf{Jumpei Ukita\textsuperscript{1,2}},
  \textbf{Yusuke Iwasawa\textsuperscript{1,2}},
  \textbf{Yutaka Matsuo\textsuperscript{1,2}}
\\
  \textsuperscript{1}Matsuo Institute, Inc., Japan,\quad
  \textsuperscript{2}The University of Tokyo, Japan
\\
  \texttt{kiyoshi.kurihara@matsuo-institute.com}
}

\begin{document}
\maketitle

\begin{abstract}
\noindent Most large language model (LLM)-based text-to-speech (TTS) methods offer no explicit function for correcting pronunciation. Consequently, these systems frequently mispronounce many everyday words, such as proper nouns, heteronyms, numerals, and domain-specific terms. While weight-editing or fine-tuning can mitigate this, these solutions risk quality degradation and must operate offline. We propose SpeechSplice, a training-free encoder-output replacement approach for pronunciation control in encoder--decoder LLM-based TTS. SpeechSplice replaces the encoder-output hidden states of a word with those of its intended reading. Cross-architecture and cross-lingual experiments cover ten public TTS systems in 17 model--language conditions. At the hidden state we replace, the intervention moves pronunciation toward the target reading more often when the generator reads the text through cross-attention than when text and speech share one self-attention sequence. In human evaluations, native-language listeners identified speech generated with SpeechSplice as realizing the target reading in 98.6--100.0\% of judgments, while a separate study found no significant naturalness difference from unmodified synthesis. By clarifying the conditions under which pronunciation correction succeeds, this study offers an empirical basis for investigating pronunciation control in LLM-based TTS architectures.
\end{abstract}

\section{Introduction}
\label{sec:introduction}

Recent large language model (LLM)-based text-to-speech (TTS) methods---including VALL-E \citep{chen2025valle}, CosyVoice~2 \citep{du2024cosyvoice2scalablestreaming}, Llasa \citep{ye2025llasa}, Qwen3-TTS \citep{hu2026qwen3tts}, and T5Gemma-TTS \citep{arata2026t5gemma}---integrate traditional language front ends into a single sequence model that maps text directly to speech. That design improves naturalness, multilingual coverage, and zero-shot voice cloning, but in most such systems it removes the editable pronunciation interface at inference time: in conventional pipelines \citep{shen2018tacotron2}, pronunciation ambiguity---heteronyms, polyphones, homographs---is handled by grapheme-to-phoneme (G2P) modules that can be updated entry by entry \citep{taylor2009tts,sproat1996multilingual,zhang2023kanji}. A few LLM-based systems offer pronunciation control at the phoneme or pinyin level \citep{cui2025glmtts,gong2026mosstts,deng2025indextts}. Fine-tuning can restore such control \citep{kato2025uttertunelorabasedtargetlanguagepronunciation}, but in LLM-based TTS it is reported to degrade quality, with catastrophic forgetting inducing speech collapse \citep{fujita2024noise,kwon25_interspeech,hsiao25_interspeech}.

\begin{figure}[!t]
\centering
\includegraphics[width=\columnwidth]{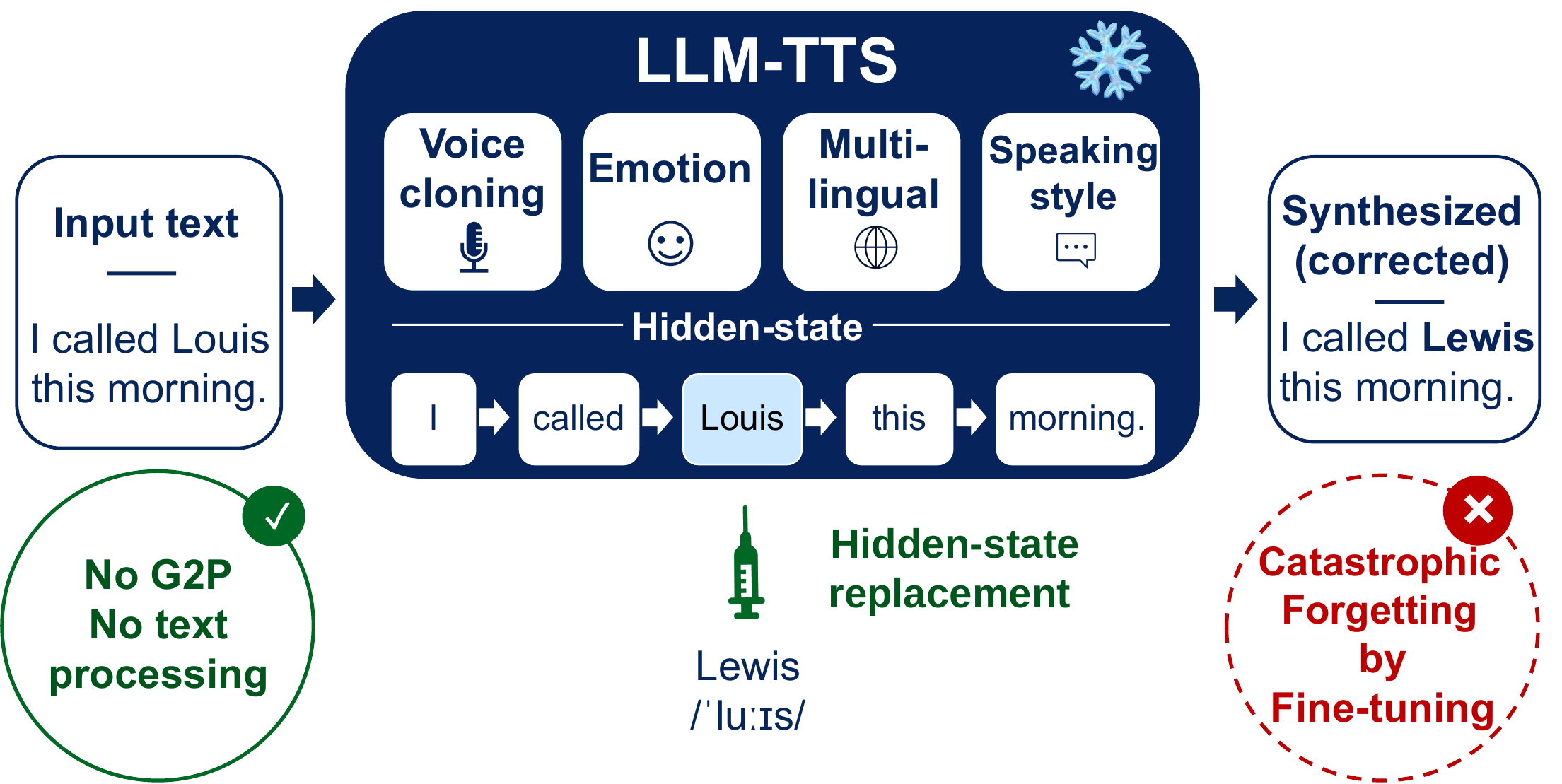}
\caption{Overview of SpeechSplice, which controls pronunciation without a grapheme-to-phoneme (G2P) front end. We evaluate preservation only for speaker similarity, naturalness, pitch, and duration.}
\label{fig:p1_overview}
\end{figure}

We propose \textbf{SpeechSplice}\footnote{Code, benchmark items, and listening-test materials: \releaseurl}, a training-free encoder-output replacement approach for G2P-free word injection in encoder--decoder LLM-based TTS. By replacing entity-position hidden states at the final output of the frozen text encoder with vectors encoding the target reading (Figure~\ref{fig:p1_overview}), it keeps pronunciation entirely within the model and gives local control without touching any weight.

\newpage
\noindent Our contributions are:

\begin{enumerate}
\item We introduce \textbf{SpeechSplice}, which injects words at the text-encoder output of encoder--decoder LLM-based TTS without a linguistic front end; all model weights stay frozen.
\item We identify which systems the intervention reaches, over ten public TTS systems in 17 model--language conditions. One requirement separates the conditions in which it works from those in which it does not: at the tensor we replace, the generator must read the encoded text through cross-attention. A second condition is presupposed rather than tested: the tokenizer must give the source and target words different token identifiers, and Section~\ref{sec:datasets} describes the case in which it does not. This finding offers an empirical basis for investigating pronunciation control in LLM-based TTS architectures.
\item We evaluate SpeechSplice with native listeners of English, Japanese, and Chinese, using paired comparison tests that include a ``neither'' option to assess target-word injection, and a mean opinion score (MOS) study to assess naturalness. Listeners select the SpeechSplice output as realizing the target reading in 98.6--100.0\% of judgments, and the MOS study indicates no significant difference from unmodified synthesis.
\end{enumerate}

\section{Related Work}
\label{sec:related}

\paragraph{Controllable speech synthesis.}
Recent work on controllable TTS organizes the field by both the target of control---such as emotion, zero-shot voice cloning, style, or pronunciation---and the underlying generation paradigm, including autoregressive codec language models, diffusion models, and hybrid systems \citep{DBLP:conf/emnlp/XieRZWL25}. Representative methods control style with Global Style Tokens \citep{wang2018gst}, control voice and speaking style with text prompts \citep{leng2024prompttts2}, and support zero-shot voice cloning \citep{chen2025valle,le2023voicebox,du2024cosyvoice2scalablestreaming}. SpeechSplice adds pronunciation control without retraining the model.

A parallel line of work improves the front end rather than removing it, through neural grapheme-to-phoneme conversion \citep{gorman2020g2p,cheng2024g2psurvey} and polyphone disambiguation \citep{zhang2022polyphonebert}. These settle a reading before synthesis and need a front end to receive it. Dictionary-free methods correct at synthesis time instead, with non-target-speaker reference speech \citep{fong22_interspeech}, voice-prompted speech editing \citep{kashyap2025prompttocorrect}, or acoustically informed spellings \citep{fong23_ssw}. FlowEdit optimizes and stores a text-embedding perturbation for each correction \citep{singh2026flowedit}. SpeechSplice needs no recording of the target word and no optimization, and it intervenes at the final encoder output.

\begin{figure}[!t]
\centering
\includegraphics[width=\columnwidth]{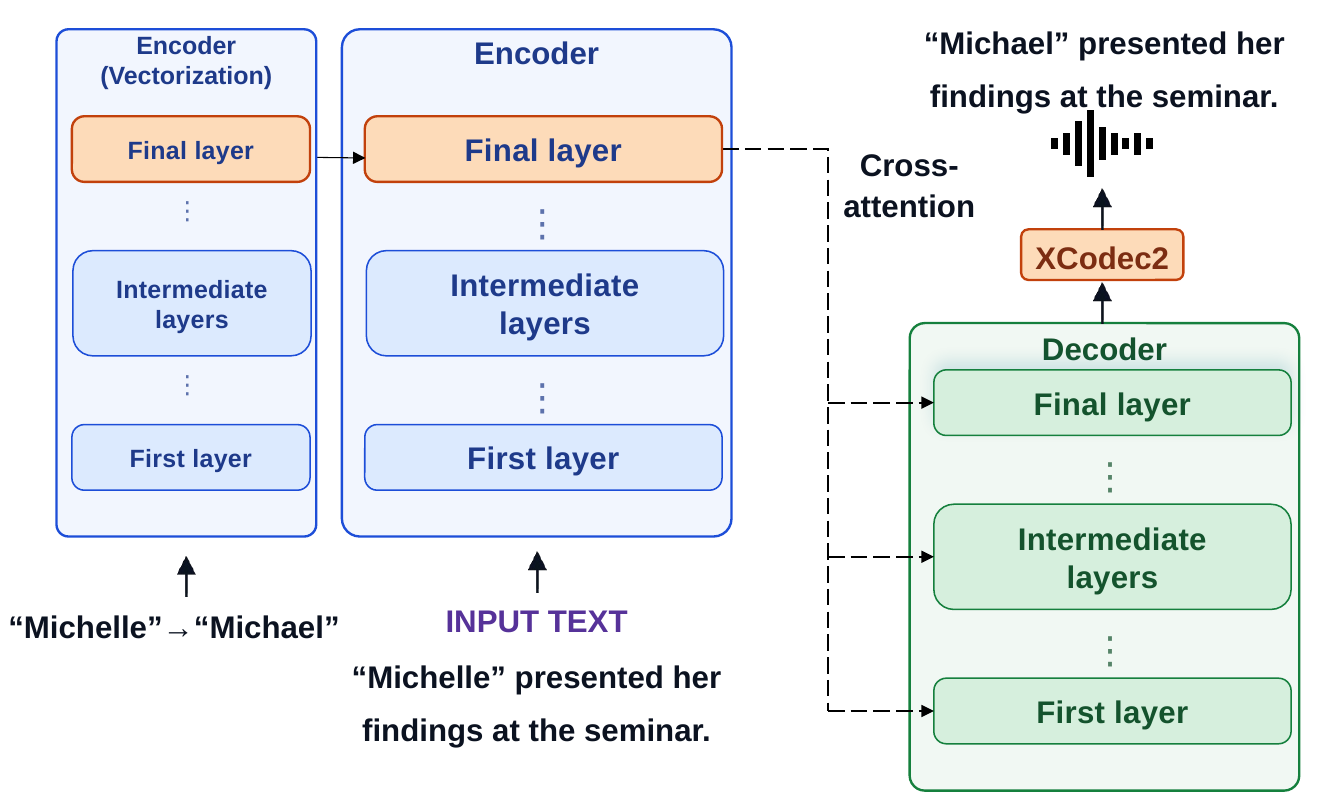}
\caption{Overview of SpeechSplice on T5Gemma-TTS. The source sentence and its target-reading version are encoded by the same frozen encoder. At the final encoder output, SpeechSplice replaces the source entity states with the corresponding target-reading states, and the decoder reads the modified sequence through cross-attention.}
\label{fig:method_overview}
\end{figure}

\paragraph{Layer-wise probing of speech representations.}
Layer-wise analyses of speech self-supervised learning (SSL) models such as wav2vec~2.0 \citep{baevski2020wav2vec2} and WavLM \citep{chen2022wavlm} show that acoustic and linguistic attributes emerge at different depths rather than uniformly \citep{pasad2021layerwise}, that phonetic information survives across many layers \citep{choi2024phonetic}, and that intermediate layers capture word-level structure \citep{pasad2024words}. Applied to question answering, logical reasoning, speech recognition, and machine translation models, letting the decoder cross-attend to intermediate encoder layers instead of the final encoder output shows that low and intermediate layers already solve simpler subtasks \citep{langedijk2024decoderlens}. While prior TTS work uses SSL layers to drive the transformation from text to phonetic information \citep{11462776,wang2023comparative,sivaguru23_interspeech,NEURIPS2022_69c754f5}, SpeechSplice takes a different approach. We apply the same layer-wise view to the text encoder of a TTS model. We then replace its final-layer states at target entity positions before the decoder reads them.

\paragraph{Activation patching and representation engineering.}
Activation patching steers model behavior at inference time by modifying internal activations rather than model weights \citep{vig2020investigating,geiger2021causal,zou2023representation,li2023iti}. In text language models, \citet{li2023iti} add learned directions to hidden states. A language model can also explain hidden states in natural language \citep{ghandeharioun2024patchscopes}, and hidden states moved from one forward pass into another can change a translated concept but not the output language \citep{dumas2025separating}. Related ideas have appeared in TTS through counterfactual activation editing \citep{lee2025counterfactual}, codec-token editing \citep{peng2024voicecraft,wang2025ssrspeech}, score-based speech editing \citep{tae2022editts}, spectrogram-level speech editing \citep{lv2026ast,jiang2023fluentspeech}, and accent editing via activation steering (\citealp{yang2026activation}; in speech recognition, \citealp{sun2026activation_asr}). Recent TTS studies also trace an emotion circuit with activation patching \citep{du2026emotioncircuit}, steer speaking rate \citep{verdini2026speakingrate}, and control speech attributes through sparse-autoencoder features \citep{koriagin2026sae}. SpeechSplice changes a word's reading by replacing its final encoder states with target-reading states. Our cross-attention ablation (Section~\ref{sec:convergence_results}) blocks attention edges, an intervention that \citet{geva2023dissecting} use to locate where information propagates to the prediction.

\paragraph{Fine-tuning-based pronunciation control.}
A separate line controls pronunciation by updating parameters, through low-rank adaptation \citep{hu2022lora}, as in UtterTune \citep{kato2025uttertunelorabasedtargetlanguagepronunciation}, and SonoEdit \citep{li2025sonoedit} edits weights in decoder-only LLM-based TTS. As noted in Section~\ref{sec:introduction}, fine-tuning LLM-based TTS is reported to degrade quality \citep{fujita2024noise,kwon25_interspeech,hsiao25_interspeech}. SpeechSplice leaves all parameters frozen instead.

\section{Method}
\label{sec:methods}
\label{sec:architecture}

\subsection{Task Definition}
\label{sec:task_definition}

Our goal is to modify only the pronunciation of a target entity at inference time while keeping both the model weights and the rest of the utterance unchanged. Given a frozen encoder--decoder TTS model, an input sentence $x$ that contains an ambiguous entity $e$, and a desired reading $r_e$ supplied at inference time, we generate speech $\tilde{a}$ in which $e$ is spoken as $r_e$ while speaker characteristics and overall speech quality stay unchanged.

Let $x'$ be the target sentence, that is, $x$ with $e$ replaced by a written form that carries the reading $r_e$. Let $\mathcal{I}_{\mathrm{src}}(e)=(s_1,\ldots,s_m)$ and $\mathcal{I}_{\mathrm{tgt}}(e)=(u_1,\ldots,u_n)$ be the entity token spans in $x$ and in $x'$. To find $\mathcal{I}_{\mathrm{src}}(e)$, we tokenize the whole sentence with character offsets and take every token whose offset range overlaps the entity. We encode $x$ and $x'$ with the same frozen TTS encoder to obtain $\mathbf{H}^{\mathrm{src}}$ and $\mathbf{H}^{\mathrm{tgt}}$. We train no parameters, so SpeechSplice is training-free. It splices the target entity states between the source prefix and suffix (Figure~\ref{fig:method_overview}):
\begin{equation}
\tilde{\mathbf{H}} = \big(\mathbf{h}^{\mathrm{src}}_{1:s_1-1},\ \mathbf{h}^{\mathrm{tgt}}_{u_1:u_n},\ \mathbf{h}^{\mathrm{src}}_{s_m+1:T}\big),
\label{eq:splice}
\end{equation}
where $\mathbf{h}_{a:b}$ is the block of states from index $a$ to $b$ and $T$ is the encoder sequence length of $x$. The splice changes the encoder length from $T$ to $T-m+n$, and we rebuild the attention mask and the position indices to match. For $m=n$ it is a replacement in place. When the length has to stay fixed we pool instead, using $A^e \in \mathbb{R}^{m \times n}$ with $A^e_{jk} \ge 0$ and $\sum_{k=1}^{n} A^e_{jk}=1$:
\begin{equation}
\bar{\mathbf{h}}_{s_j}^{\mathrm{tgt}}
= \sum_{k=1}^{n} A^e_{jk}\mathbf{h}_{u_k}^{\mathrm{tgt}},
\label{eq:aligned_target}
\end{equation}
and we write $\bar{\mathbf{h}}^{\mathrm{tgt}}_t$ at each $t \in \mathcal{I}_{\mathrm{src}}(e)$. The ablation of Section~\ref{sec:convergence_results} takes this form with $A^e_{jk}=1/n$.

\subsection{The Decoder-Facing Encoder Output}
\label{sec:base_model}

In encoder--decoder TTS, the encoder maps the input text to $\mathbf{H} = [\mathbf{h}_1, \ldots, \mathbf{h}_T] \in \mathbb{R}^{T \times d}$, where $T$ is the encoder sequence length and $d$ is the hidden-state dimensionality. The decoder generates acoustic or codec tokens while reading this sequence through cross-attention. The final encoder output is therefore the text-side interface---and the bottleneck---between text encoding and speech generation. Retrieval and injection both take place at the final encoder output. SpeechSplice reads the target entity states there and splices them into the source sequence at the entity-token positions, as in Eq.~\eqref{eq:splice}. Appendix~\ref{app:compute} specifies the text-side tensor that we replace in each of the other nine systems compared in Section~\ref{sec:mechanism_results}.

\section{Experimental Setup}
\label{sec:experimental_setup}

\subsection{Datasets and Models}
\label{sec:datasets}

\paragraph{TTS models.}
We evaluated ten public TTS systems as frozen checkpoints (links in Appendix~\ref{app:compute}). In five, the decoder reads the text encoder output through cross-attention (encoder--decoder): T5Gemma-TTS, LongCat-AudioDiT \citep{xin2026longcat}, Irodori-TTS \citep{arata2026irodori}, Kiseki-TTS \citep{kiseki2026}, and SpeechT5 \citep{ao2022speecht5}. In the other five, the text shares one self-attention sequence with the speech tokens (decoder-only): Qwen3-TTS, XTTS-v2 \citep{casanova2024xtts}, Llasa-3B \citep{ye2025llasa}, CSM-1B, and Bark. T5Gemma-TTS and Qwen3-TTS are evaluated in English (EN), Japanese (JA), and Chinese (ZH), and each of the others in one or two of these languages, giving 17 model--language conditions. SpeechT5 is evaluated in EN only: its English-only tokenizer gives the source and target words of every JA and ZH pair the same token identifiers, so replacement cannot change the encoder output, and we exclude those two conditions rather than count them as failures. The method therefore presupposes a tokenizer that gives the source and target words different token identifiers.

\paragraph{Benchmarks.}
The cross-lingual benchmark has 300 pairs per language, built from 50 hand-written entity pairs and six sentence templates. The cross-architecture benchmark, which the acoustic cross-attention ablation and Appendix~\ref{app:pronunciation_baselines} also use, has another 300 pairs per language, built from 30 hand-written entity pairs and ten templates, with a Japanese surname as the target and an English given name (EN) or a Chinese surname (JA, ZH) as the source. Source and target sentences differ in exactly one entity span.

\paragraph{Conventional system comparison.}
The other-system comparison used the English split of the cross-lingual benchmark. We compared SpeechSplice against VoiceCraft zero-shot TTS \citep{peng2024voicecraft} and EdiTTS content replacement \citep{tae2022editts}. SpeechSplice and VoiceCraft used the same 3.25-s LibriTTS \citep{zen2019libritts} test-clean voice prompt; EdiTTS could not clone an arbitrary prompt speaker and instead edited speech synthesized by its Grad-TTS \citep{popov2021gradtts} backbone, trained on LJSpeech\urlfootnote{LJ Speech dataset}{https://keithito.com/LJ-Speech-Dataset/}.

\paragraph{Human-listening stimuli.}
We built stimuli for two studies. Baseline (BL) and injected (INJ) clips were generated with T5Gemma-TTS voice cloning from one native reference speaker per language, taken from LibriTTS (EN), JSUT \citep{sonobe2017jsut} (JA), and the Databaker Chinese Standard Mandarin Speech Corpus\urlfootnote{Databaker Chinese Standard Mandarin Speech Corpus}{https://www.data-baker.com/open_source.html} (ZH), using the variable-length splice implementation. The paired comparison test drew 50 name items per language from the cross-lingual benchmark, except one JA item that was replaced before data collection, and additionally used a Japanese same-surface set: 50 sentences in which the frozen T5Gemma-TTS baseline misreads the target word while the written sentence stays unchanged. These items cover heteronyms, numerals, okurigana, place names, and rare surnames; Appendix~\ref{app:same_surface_human} gives their sources and composition. The naturalness study used another 50 baseline--injection name-replacement pairs in each of EN, JA, and ZH, built and synthesized separately from the 300 pairs and the paired-comparison items. BL and INJ used the same text, voice prompt, and decoding configuration, and the random seed was reset to the same value before every synthesis call, so the two conditions differ only in the encoder-output splice intervention.

\subsection{Evaluation Metrics}
\label{sec:eval_metrics}

\paragraph{Automatic metrics.}
To measure whether injected speech carries the target reading and whether the rest of the speech is preserved, we use the following metrics. \textbf{Whisper-INJ} is the percentage of injected samples whose normalized automatic speech recognition (ASR) transcript contains the target reading. Transcripts come from Whisper large-v3 \citep{radford2023robust}; on the cross-lingual benchmark and in the two-entity test, JA uses Furigana Whisper instead and ZH is matched after pinyin normalization. Appendix~\ref{app:cer} reports the full-sentence character error rate (CER). UTMOS \citep{saeki2022utmos} measures speech quality. Speaker similarity is the cosine similarity of speaker-verification embeddings from WavLM-SV \citep{chen2022wavlm} between each baseline clip and its injected or edited counterpart, and between each injected clip and the voice prompt. The $|\Delta F_0|$ ratio is the mean absolute change in fundamental frequency $F_0$ from the baseline divided by the mean change between two baseline utterances of different sentences, and $|\Delta\,\mathrm{duration}|$ is the mean absolute change in utterance duration from the baseline.

\paragraph{Paired comparison test.}
To verify that the injected target word was correctly realized in the generated speech, we conducted a paired comparison test following the pronunciation-evaluation protocols of Speech Audio Corrector \citep{fong22_interspeech} and Spell4TTS \citep{fong23_ssw}: native listeners heard a BL--INJ pair in randomized order and judged which clip realized a displayed target word, with a third option of choosing neither clip. Whereas \citet{fong22_interspeech,fong23_ssw} present the intended pronunciation as reference audio or an alternative spelling, we display the target word itself, since for proper names neither clip is self-evidently correct without a reference; EN and ZH displayed the orthographic form, and JA the kana reading alone. Each language used 50 items in two 25-item blocks; twenty participants per language (ten per block) each judged their 25 items plus three catch trials, for ten independent judgments per item and 500 per language. The Japanese same-surface set followed the same protocol (Appendix~\ref{app:same_surface_human}). The interface disabled answering until both clips had played to the end, and catch trials displayed a word spoken in neither clip. We report the unconditional INJ-selection rate, counting a choice of neither clip as failure, with item-level bootstrap confidence intervals. The name-set clips are drawn from the cross-lingual benchmark, so for each item we also compare the majority of its ten judgments with the Whisper judgment of the same pair; one replaced Japanese item lacks an automatic judgment. No participant met a pre-specified exclusion criterion: failing more than one catch trial, straight-line responding, or completing fewer than 90\% of the trials.

\paragraph{Naturalness MOS.}
A separate study measured whether injection degrades naturalness. Twenty native-speaker participants per language rated the 50 BL and 50 INJ clips on a five-point scale, providing ten ratings per clip on average, and the instructions explicitly excluded pronunciation correctness from the rating criteria. BL--INJ differences were tested with two-sided paired Wilcoxon signed-rank tests over item means. Appendix~\ref{app:human_eval_details} documents recruitment, payment, consent, the instructions, and the interface of both studies.

\paragraph{Cross-architecture metrics.}
To compare systems on how far injection moves the speech and how often it moves the entity pronunciation toward the target reading, we report whole-utterance BL--INJ mel-cepstral distortion (MCD) \citep{kubichek1993mel} and Phoneme Injection-Better (PIB) with Wilson 95\% confidence intervals (CIs). PIB is the fraction of the 300 pairs for which injection lowers the entity phoneme error rate (PER) measured against the target reading,
\begin{equation}
\mathrm{PIB} = \frac{1}{N}\sum_{p=1}^{N} \mathbf{1}[\,\mathrm{PER}^{\mathrm{INJ}}_p < \mathrm{PER}^{\mathrm{BL}}_p\,].
\label{eq:phoneme_better}
\end{equation}
We write $L$ for the number of layers the text passes through in a system and $l^{*}$ for the number of those layers that precede the replacement, so our replacement sits at $l^{*}{=}L$ in the encoder--decoder systems and at $l^{*}{=}L{-}1$ in the decoder-only ones; Section~\ref{sec:layer_results} sweeps $l^{*}$ with everything else fixed.

\paragraph{Multi-entity test.}
To examine whether two entities in one sentence can be corrected together, we used a further 300 sentences per language that carry two entity slots each, built from 50 pairs of same-kind proper nouns and six templates of their own. We corrected one entity, then both, and scored Whisper-INJ per slot, pooling 600 slot judgments from 300 sentences; the two slots of a sentence are not independent, so we compare the columns by resampling sentences (Appendix~\ref{app:multientity_quality}). The BL column scores the baseline speech for the words it was asked to read.

\paragraph{Cross-attention ablation.}
To examine whether entity-position cross-attention carries the replaced states to the generated speech, we also synthesized BL and INJ with entity-position cross-attention blocked (BL$_{\mathrm{mask}}$, INJ$_{\mathrm{mask}}$) and compared each injected condition with its matched baseline over the entity span, by MEL Similarity and by a variance-normalized MCD whose values are comparable only within Table~\ref{tab:crossattn_mask} (Appendix~\ref{app:crossattn_acoustic}).

\setcounter{table}{0}
\begin{table*}[!t]
\centering
\footnotesize
\setlength{\tabcolsep}{2pt}
\resizebox{\textwidth}{!}{%
\begin{tabular}{@{}cccccccccccc@{}}
\toprule
& \multicolumn{2}{c}{Paired comparison test} & \multicolumn{3}{c}{Naturalness MOS}
& & \multicolumn{2}{c}{UTMOS} & \multicolumn{2}{c}{Speaker similarity} & \\
\cmidrule(lr){2-3}\cmidrule(lr){4-6}\cmidrule(lr){8-9}\cmidrule(lr){10-11}
Lang. & \makecell{INJ selected\\(\%)~$\uparrow$} & $z$ ($p$) & BL & INJ~$\uparrow$ & $p$
& \makecell{Whisper-INJ\\(\%)~$\uparrow$} & BL & INJ~$\uparrow$
& BL--INJ~$\uparrow$ & INJ--prompt~$\uparrow$
& \makecell{$|\Delta F_0|$\\ratio~$\downarrow$} \\
\midrule
EN & $99.4~(+0.6/{-}0.8)$ & \makecell{$9.07$\\($<10^{-16}$)} & $3.56 \pm 0.09$ & $3.52 \pm 0.08$ & $0.45$ & \makecell{$96.33$\\$(+1.61/{-}2.78)$} & $3.86 \pm 0.05$ & $3.82 \pm 0.05$ & $0.83 \pm 0.01$ & $0.84 \pm 0.01$ & $0.98$ \\
JA & $98.6~(+1.4/{-}2.8)$ & \makecell{$4.24$\\($2.3\times10^{-5}$)} & $3.68 \pm 0.09$ & $3.64 \pm 0.09$ & $0.74$ & \makecell{$92.33$\\$(+2.50/{-}3.57)$} & $3.39 \pm 0.04$ & $3.41 \pm 0.05$ & $0.94 \pm 0.00$ & $0.85 \pm 0.01$ & $0.71$ \\
ZH & $100.0~(+0.0/{-}0.0)$ & \makecell{--\\($1.8\times10^{-15}$)} & $3.67 \pm 0.10$ & $3.71 \pm 0.10$ & $0.96$ & \makecell{$96.00$\\$(+1.70/{-}2.86)$} & $3.33 \pm 0.07$ & $3.35 \pm 0.07$ & $0.91 \pm 0.01$ & $0.89 \pm 0.01$ & $0.88$ \\
\bottomrule
\end{tabular}%
}
\caption{Participants chose the injected clip in almost every judgment, and MOS showed no significant BL--INJ difference. We did not test for equivalence. BL: T5Gemma-TTS baseline synthesis; INJ: SpeechSplice injection. $z$ ($p$): binomial generalized estimating equation (GEE) test against chance \citep{liang1986longitudinal}, or an exact binomial test for ZH (Appendix~\ref{app:he_stats}). MOS $p$: two-sided paired Wilcoxon signed-rank test over item means. Speaker similarity: WavLM-SV cosine similarity between the two clips named (prompt: voice prompt). 95\% intervals: item-level bootstrap for INJ selected and Wilson for Whisper-INJ, as $(+\mathrm{upper}/{-}\mathrm{lower})$; $\pm 1.96\times$ standard error for means.}
\label{tab:cross_lingual_overview}
\end{table*}

\begin{table*}[!t]
\centering
\small
\setlength{\tabcolsep}{1.2pt}
\begin{tabular}{@{}lcccccccc@{}}
\toprule
& & \multicolumn{3}{c}{UTMOS} & \multicolumn{2}{c}{Speaker similarity} & & \\
\cmidrule(lr){3-5}\cmidrule(lr){6-7}
System & \makecell{Whisper-INJ\\(\%)~$\uparrow$} & BL & \makecell{Injected/\\edited~$\uparrow$} & $p$ & BL--INJ~$\uparrow$ & INJ--prompt~$\uparrow$ & \makecell{$|\Delta F_0|$\\ratio~$\downarrow$} & \makecell{$|\Delta\,\mathrm{duration}|$\\(s)~$\downarrow$} \\
\midrule
SpeechSplice (ours) & $96.33~(+1.61/{-}2.78)$ & $3.86 \pm 0.05$ & $3.82 \pm 0.05$ & $0.07$ & $0.83 \pm 0.01$ & $0.84 \pm 0.01$ & $0.98$ & $0.00 \pm 0.00$ \\
VoiceCraft (zero-shot) & $90.00~(+2.91/{-}3.92)$ & $3.62 \pm 0.04$ & $3.63 \pm 0.04$ & $0.54$ & $0.87 \pm 0.01$ & $0.87 \pm 0.01$ & $0.98$ & $0.20 \pm 0.02$ \\
EdiTTS$^{\dagger}$ & $86.00~(+3.47/{-}4.38)$ & $4.29 \pm 0.02$ & $4.28 \pm 0.02$ & $0.09$ & $0.97 \pm 0.00$ & -- & $0.74$ & $0.13 \pm 0.01$ \\
\bottomrule
\end{tabular}
\caption{Comparison with conventional methods on the English split of the cross-lingual benchmark. SpeechSplice recovers the target reading most often and keeps the BL duration by construction. The SpeechSplice row is the EN condition of Table~\ref{tab:cross_lingual_overview}, whose caption gives the intervals; Section~\ref{sec:eval_metrics} defines the columns. UTMOS $p$: two-sided paired Wilcoxon signed-rank test over per-clip scores. VoiceCraft (zero-shot): zero-shot voice cloning. BL: synthesis of the source sentence (Grad-TTS for EdiTTS); Injected/edited (INJ): SpeechSplice, VoiceCraft target-sentence synthesis, or the EdiTTS edit. $^{\dagger}$EdiTTS is not LLM-based and cannot clone a voice, so INJ--prompt is unavailable.}
\label{tab:system_comparison}
\end{table*}

\section{Results}
\label{sec:results}

\subsection{Cross-Lingual Injection and Speech Quality}
\label{sec:cross_lingual_divergence_results}
\label{sec:rtf_quality}

Listeners confirmed the injected reading directly
(Table~\ref{tab:cross_lingual_overview}): in the name sets no participant
chose the baseline, consistent with the Whisper-INJ scores. Full-sentence CER shows the same shift: the injected transcripts lie close to the target sentence and far from the source sentence in all three languages (Appendix~\ref{app:cer}). The Japanese
same-surface set, which corrects naturally occurring misreadings rather than
replacing words, gave the same result: only a few of its judgments went to the baseline or to neither clip (Table~\ref{tab:same_surface_results}). The selection rates significantly exceed chance in every set, under the generalized estimating equation (GEE) test or, for ZH, an exact binomial test. Automatic scoring of a 160-item Japanese reading benchmark agrees: injection corrected all 45 strict baseline misreadings (Appendix~\ref{app:ja_reading_benchmark}). MOS
showed no significant BL--INJ difference in any language and UTMOS no
consistent change; speaker similarity to the voice prompt did not move in any language in paired
Wilcoxon tests over the WavLM-SV cosine similarities of the 300 automatic-evaluation pairs,
while BL--INJ similarity stayed high; and every
$|\Delta F_0|$ ratio falls below 1, so injection shifts $F_0$ less than
ordinary sentence-to-sentence variation does. Listener majorities matched the Whisper judgments on all but eight name items, and the ASR was the stricter judge in seven of the eight disagreements (Appendix~\ref{app:failures}), so the automatic scores are conservative relative to human perception.

\subsection{Comparison with Conventional Methods}
\label{sec:baseline_comparison_results}

We compared SpeechSplice with VoiceCraft zero-shot TTS and EdiTTS on the
English split of the cross-lingual benchmark (Table~\ref{tab:system_comparison}), to
assess whether hidden-state injection recovers the target reading more often than
existing generation and editing approaches.

SpeechSplice recovered the target reading more often than VoiceCraft, which
resynthesizes the whole utterance, and than EdiTTS, which replaces content by
diffusion. The
three systems synthesize in different acoustic domains, so we read UTMOS within
each system rather than across them; none showed a significant change from its
own baseline, so we detected no quality cost for any of the three. For speaker
identity we tested the change against the voice prompt directly: the cosine similarity
moves by $-0.001$ for SpeechSplice, with a 95\% CI of $\pm 0.009$ and $p=0.98$
in a paired Wilcoxon test over the 300 pairs, and by $-0.009$ for VoiceCraft, with a
95\% CI of $\pm 0.008$ and $p=0.026$,
the only drift that reaches significance, while EdiTTS cannot clone a prompt
speaker, so this test does not apply to it. All three
keep the pitch change small, and EdiTTS shifts pitch least. The interfaces differ most in
duration: SpeechSplice reuses the BL generation length by construction (Appendix~\ref{app:compute}), while
VoiceCraft and EdiTTS change it. SpeechSplice thus
gained accuracy at the BL length, and we
detected no cost elsewhere.

\setcounter{table}{2}
\begin{table*}[!t]
\centering
{\footnotesize
\setlength{\tabcolsep}{1.7pt}
\renewcommand{\arraystretch}{1.04}
\begin{tabular}{llcccccccc}
\toprule
& & & \multicolumn{2}{c}{At our replacement point} & \multicolumn{2}{c}{Layer sweep} & \multicolumn{3}{c}{Two-entity sentences (\%)} \\
\cmidrule(lr){4-5}\cmidrule(lr){6-7}\cmidrule(lr){8-10}
Model & Lang. & $L$ & MCD~(dB) & PIB~(\%)~$\uparrow$ & $l^{*}$ & \makecell{PIB~(\%)\\min--max} & BL & \makecell{INJ\\(one)~$\uparrow$} & \makecell{INJ\\(both)~$\uparrow$} \\
\midrule
\multicolumn{10}{l}{\emph{Encoder--decoder (cross-attention)}} \\
\multirow{3}{*}{T5Gemma-TTS} & EN & 26 & $9.95 \pm 0.20$ & $57.00~(+5.48/{-}5.66)$ & 1/8/15/22/26 & 56.67--60.00 & 97.17 & 97.17 & 96.67 \\
 & JA & 26 & $8.24 \pm 0.21$ & $52.00~(+5.59/{-}5.64)$ & 1/8/15/22/26 & 48.00--52.00 & 88.00 & 88.83 & 89.17 \\
 & ZH & 26 & $8.96 \pm 0.25$ & $55.00~(+5.53/{-}5.66)$ & 1/8/15/22/26 & 52.33--55.67 & 78.83 & 81.83 & 81.00 \\
\multirow{2}{*}{LongCat-AudioDiT} & EN & 12 & $10.00 \pm 0.24$ & $65.33~(+5.16/{-}5.55)$ & 1/3/6/9/12 & 59.33--65.33 & 94.67 & 96.33 & 95.17 \\
 & ZH & 12 & $8.65 \pm 0.25$ & $62.67~(+5.28/{-}5.60)$ & 1/3/6/9/12 & 58.33--64.67 & 99.83 & 99.33 & 99.17 \\
Irodori-TTS & JA & 25 & $7.20 \pm 0.24$ & $49.33~(+5.63/{-}5.61)$ & 1/7/13/19/25 & 49.33--59.33 & 96.33 & 94.33 & 94.67 \\
Kiseki-TTS & JA & 12 & $5.28 \pm 0.17$ & $41.67~(+5.65/{-}5.44)$ & 1/3/6/9/12 & 41.67--44.67 & \phantom{0}2.67 & \phantom{0}7.83 & \phantom{0}7.00 \\
SpeechT5 & EN & 12 & $3.53 \pm 0.06$ & $64.00~(+5.22/{-}5.58)$ & 1/3/6/9/12 & 54.00--64.00 & 95.67 & 89.33 & 90.67 \\
\midrule
\multicolumn{10}{l}{\emph{Decoder-only (self-attention)}} \\
\multirow{3}{*}{Qwen3-TTS} & EN & 28 & $2.30 \pm 0.28$ & $13.67~(+4.35/{-}3.43)$ & 1/10/14/21/27 & 12.33--15.00 & 97.83 & \phantom{0}0.00 & \phantom{0}0.00 \\
 & JA & 28 & $1.42 \pm 0.27$ & $12.00~(+4.17/{-}3.20)$ & 1/10/14/21/27 & 10.00--13.67 & 91.83 & \phantom{0}0.00 & \phantom{0}0.00 \\
 & ZH & 28 & $0.86 \pm 0.15$ & $12.33~(+4.20/{-}3.25)$ & 1/10/14/21/27 & \phantom{0}9.00--12.67 & 99.83 & \phantom{0}0.00 & \phantom{0}0.00 \\
XTTS-v2 & EN & 30 & $0.34 \pm 0.08$ & $\phantom{0}2.67~(+2.51/{-}1.31)$ & 1/9/16/24/29 & \phantom{0}2.67--54.67 & 96.33 & \phantom{0}0.00 & \phantom{0}0.00 \\
\multirow{2}{*}{Llasa-3B} & EN & 28 & $1.91 \pm 0.30$ & $13.33~(+4.31/{-}3.39)$ & 1/8/15/22/27 & 13.33--51.33 & 96.00 & \phantom{0}0.00 & \phantom{0}0.00 \\
 & ZH & 28 & $2.21 \pm 0.28$ & $20.33~(+4.92/{-}4.17)$ & 1/8/15/22/27 & 20.33--49.00 & 97.17 & \phantom{0}0.00 & \phantom{0}0.00 \\
CSM-1B & EN & 16 & $4.74 \pm 0.15$ & $27.00~(+5.29/{-}4.71)$ & 1/4/8/11/15 & 27.00--51.67 & 96.67 & \phantom{0}0.00 & \phantom{0}0.00 \\
\multirow{2}{*}{Bark} & EN & 24 & $0.58 \pm 0.18$ & $\phantom{0}4.00~(+2.86/{-}1.70)$ & 1/6/12/17/23 & \phantom{0}4.00--49.67 & 97.00 & \phantom{0}0.00 & \phantom{0}0.00 \\
 & JA & 24 & $0.97 \pm 0.26$ & $\phantom{0}7.67~(+3.57/{-}2.50)$ & 1/6/12/17/23 & \phantom{0}7.67--41.67 & 57.33 & \phantom{0}0.00 & \phantom{0}0.00 \\
\bottomrule
\end{tabular}}
\caption{PIB at our replacement point ($l^{*}{=}L$; $L{-}1$ if decoder-only) separates encoder--decoder from decoder-only conditions; most decoder-only ones rise at shallower $l^{*}$, and a second entity costs no detectable accuracy. INJ replaces an entity's hidden states at $l^{*}$ ($L$: layers the text passes through) with the swapped sentence's; BL does not. MCD: BL--INJ mel-cepstral distortion. PIB: \% of pairs where INJ lowers the entity phoneme error rate against the swapped sentence. 95\% CIs: $\pm$ for MCD; Wilson, as $(+\mathrm{upper}/{-}\mathrm{lower})$, for PIB. Two-entity sentences: \% of 600 slots read as intended (BL: original; INJ: one or both swapped in).}
\label{tab:cross_arch_main}
\end{table*}

\subsection{Cross-Architecture Results}
\label{sec:mechanism_results}

This section examines how far the intervention carries beyond T5Gemma-TTS and what separates the systems on which it works from those on which it does not. Table~\ref{tab:cross_arch_main} reports this comparison for all 17 model--language conditions, together with the layer sweep of Section~\ref{sec:layer_results} and the two-entity test of Section~\ref{sec:multientity_results}. Across them the split follows the path by which the generator reads the encoded text. At the tensor we replace, the PIB values of the two groups do not overlap: the lowest encoder--decoder value, Kiseki-TTS JA, lies above the highest decoder-only value, CSM-1B EN. The other three encoder--decoder systems bring two decoder families that T5Gemma-TTS and SpeechT5 do not use---a diffusion transformer \citep{peebles2023dit} in LongCat-AudioDiT and in Irodori-TTS, and a Mamba-2 decoder \citep{dao2024mamba2} in Kiseki-TTS---so the pattern is not specific to the autoregressive decoder those two share. MCD is above zero in every condition, including the decoder-only ones, where even the smallest value, that of XTTS-v2 EN, has an interval that excludes zero; there the replacement changes the speech but, by PIB, moves it toward the target reading less often than in any encoder--decoder condition.

Bark EN has the second-lowest PIB in the table: in the large majority of its pairs the injected semantic tokens came out identical to the baseline, so the PER cannot change and the pair counts as a failure by definition.

\subsection{Effect of Replacement Depth}
\label{sec:layer_results}

Depth has little effect in the encoder--decoder systems: in Table~\ref{tab:cross_arch_main},
the PIB of T5Gemma-TTS EN varies across the whole encoder by less than the
width of the confidence interval of any single point, and SpeechT5, LongCat-AudioDiT, Irodori-TTS, and Kiseki-TTS behave the same way,
consistent with the decoder reading one tensor that an earlier replacement
still reaches through the remaining layers. Depth has a large effect in the decoder-only
systems: CSM-1B EN,
Llasa-3B EN and ZH, Bark EN and JA, and XTTS-v2 EN all reach a much higher PIB at
$l^{*}{=}1$ than at our replacement point. There the speech tokens attend to the text rows
through the same stack, so a replacement at depth $l^{*}$ rewrites the keys and
values of every layer above it. Qwen3-TTS shows this is not universal: its PIB
stays low at every swept depth.

\subsection{Multi-Entity Correction}
\label{sec:multientity_results}

Correcting both entities has no detectable cost relative to correcting one: across the eight
encoder--decoder conditions in Table~\ref{tab:cross_arch_main}, the two columns
differ in both directions, and the 95\% interval of their paired per-sentence difference
includes zero in every condition (Appendix~\ref{app:multientity_quality}). The decoder-only conditions read the corrected word in no slot
at all, although their baseline speech reads the original words in the vast majority of slots, and in over half of them for Bark JA, which suggests that the failure lies in the transmission rather than in reading the stimuli. Replacing both entities also leaves UTMOS close to the baseline in every condition (Appendix~\ref{app:multientity_quality}).

\subsection{Cross-Attention Ablation}
\label{sec:convergence_results}

Section~\ref{sec:mechanism_results} suggests that decoder-to-encoder cross-attention mediates how entity-position encoder states affect speech generation. We conducted the four-condition T5Gemma-TTS ablation of Section~\ref{sec:eval_metrics} with length-preserving injection: we mean-pooled target entity states across the source span so that the encoder length stayed fixed, and we masked only cross-attention to source entity columns.

Table~\ref{tab:crossattn_mask} (Appendix~\ref{app:crossattn_acoustic}) reports the results: with cross-attention intact, injection causes large acoustic differences in all three languages; when entity-position cross-attention is blocked, the outputs become nearly identical. A token-count-matched JA subset gives the same result (Appendix~\ref{app:ja_token_subset}). These metrics measure BL--INJ acoustic divergence rather than target-direction accuracy. Appendix~\ref{app:ja_crossattn_phoneme} repeats the blocking at the phoneme level on same-surface pairs.

\section{Discussion}
\label{sec:discussion}

\paragraph{Human evaluation.}
The two listening studies answer separate questions: listeners hear the intended reading at near-ceiling rates in all
three languages and on the same-surface set, and the MOS study finds no
significant naturalness difference. Where listener majorities and the ASR
judgments disagreed, the ASR usually rejected a reading that listeners
accepted, so the automatic scores understate rather than overstate the effect.

\paragraph{Cross-architecture comparison.}
In the systems we tested, the split between the encoder--decoder and decoder-only groups (Table~\ref{tab:cross_arch_main}) is consistent with one account, built on one requirement on the generator: that it read the text side through the replaced state. The replacement changes the pronunciation when this holds, and much less often when most of that reading bypasses the replaced state. The encoder--decoder systems meet the requirement at the tensor we replace, which is the text-side interface their decoders read through cross-attention (Section~\ref{sec:base_model}; Appendix~\ref{app:compute}). The five decoder-only systems show what happens when most of that reading bypasses the replaced state: at our replacement point they change the speech---their BL--INJ MCD is nonzero in every condition---yet bring the entity closer to the target reading less often than any encoder--decoder condition, so in these systems, at that depth, a replaced text state reached the generated speech less reliably through one self-attention sequence than through a cross-attention path. The account also presupposes a tokenizer that keeps the two written forms apart: they must receive different token identifiers before any hidden state can distinguish them, which is why SpeechT5 enters in English only (Section~\ref{sec:datasets}).

\paragraph{Replacement depth.}
The layer sweep qualifies this contrast: in every decoder-only system except Qwen3-TTS, it suggests that the low values at our replacement point reflect the depth at which the replacement enters rather than the absence of any path by which the replaced state can be read. In the encoder--decoder systems an earlier replacement still reaches the tensor the decoder reads (Section~\ref{sec:layer_results}); in the decoder-only systems the audio tokens read the text rows in every layer, so with the replacement at $l^{*}{=}L{-}1$, only the last layer reads the replaced state and every earlier layer still reads the original word. At the best depth we found (Limitations), XTTS-v2, Llasa-3B, CSM-1B, and Bark reach PIB within the range of the encoder--decoder conditions at our replacement point, as Table~\ref{tab:cross_arch_main} shows. One result falls outside this account: Qwen3-TTS, which no swept depth lifts into that range.

\paragraph{Correcting two entities.}
The two-entity test is consistent with the same account. The decoder-only conditions read the corrected word in no slot, in either the INJ (one) or the INJ (both) column of Table~\ref{tab:cross_arch_main}, at our replacement point (Appendix~\ref{app:compute}), which reproduces the split with another metric and another sentence set and suggests that it is not specific to PIB or to the single-entity pairs. In the encoder--decoder conditions, correcting a second entity does not detectably lower per-slot Whisper-INJ (Section~\ref{sec:multientity_results}), consistent with each replaced span being read through cross-attention to its own positions, which the ablation of Section~\ref{sec:convergence_results} supports for T5Gemma-TTS. Their INJ (one) and INJ (both) columns also stay close to the BL column, which runs from Kiseki-TTS JA, whose baseline speech rarely reads the words it was asked to read, to LongCat-AudioDiT ZH, whose baseline speech almost always does. This suggests that where the generator reads the text side through the replaced state, the corrected rate tracks how reliably each system reads words of this kind, whether one entity is replaced or two.

\paragraph{Transmission and comparison with conventional methods.}
The ablation identifies entity-position cross-attention as the main acoustic transmission path: blocking it removes the BL--INJ difference in EN, JA, and ZH. On the controlled English benchmark SpeechSplice also recovers the target reading more often than VoiceCraft and EdiTTS. That comparison spans distinct interfaces---full continuation synthesis, diffusion-based content replacement, and encoder-state splicing---yet intervening directly at the decoder-facing encoder tensor gives the highest pronunciation accuracy while keeping the BL duration and showing no detectable change in naturalness.

\section{Conclusion}
\label{sec:conclusion}

We introduced SpeechSplice, a training-free intervention that restores pronunciation correction in encoder--decoder LLM-based TTS. Native listeners picked the injected clip in almost every judgment in EN, JA, and ZH and on a Japanese set where the baseline misreads a word that stays written the same, and a separate study found no significant naturalness difference from unmodified synthesis. SpeechSplice also recovered the target reading more often than the public synthesis and editing systems we evaluated.

At the tensor we replace, the replacement succeeds when the generator reads the encoded text through cross-attention: over 17 model--language conditions the five encoder--decoder systems reach a higher PIB than the five decoder-only systems, with no overlap between the two groups. A training-free intervention can deliver the intended reading in encoder--decoder LLM-based TTS.
\paragraph{Future work.}
 Models, including decoder-only Transformers, could be trained from scratch to support injection by meeting the requirement we identify, and phoneme-level conditioning could remove the need to respell the target word.

\section*{Limitations}
\label{sec:limitations}

\begin{itemize}
\item The mechanism is compared across ten public TTS systems, but decoder architecture, training data, language coverage, tokenizer choice, and model scale still vary together, so the comparison identifies which systems the intervention reaches rather than which single factor decides it.
\item Which depth to replace at is settled empirically. In Bark, Llasa-3B, and Qwen3-TTS EN the best depth we found is the first layer rather than the one we report in Table~\ref{tab:cross_arch_main}; in XTTS-v2, CSM-1B, and Qwen3-TTS JA and ZH it is an intermediate one ($l^{*}{=}9$, $4$, and $21$, respectively). We have no criterion that selects the depth in advance.
\item The method requires the target reading to be specified correctly at inference time.
\item Variable-length splicing preserves all target entity states when source and target token counts differ, but requires consistent reconstruction of the attention mask and position indices; portability to other positional schemes remains to be tested.
\item The MOS study uses one cloned reference speaker per language.
\item Outside the listening tests, we judge pronunciation automatically with ASR and a phoneme recognizer, and ASR can miss reading errors that listeners hear \citep{luo2026asrroundtrip}. We did not evaluate pitch accent.
\item Kiseki-TTS has the lowest PIB of the encoder--decoder systems, and even its baseline speech is rarely recognized in the two-entity test (Table~\ref{tab:cross_arch_main}).
\item The two-entity test scores pronunciation. We estimated the speech quality of its outputs only with UTMOS (Appendix~\ref{app:multientity_quality}) and did not evaluate their naturalness, intelligibility, or artifacts with listeners. Long-form text, more diverse speakers, low-resource languages, closely spaced entities, and three or more entities per sentence also remain untested.
\end{itemize}

\section*{Ethics Statement}
\label{sec:ethics}
This work edits the internal representations of a pretrained TTS model to correct the pronunciation of proper nouns. The same mechanism can also create risks, so we note the main ethical points below.

\paragraph{Potential for misuse.}
The method could be used to deliberately mispronounce names. Two factors limit this risk. The user must specify the target reading explicitly, and the intervention stays local to the chosen entity span.

\paragraph{Speaker and demographic considerations.}
More diverse evaluation is needed before deployment across different name origins and languages.

\paragraph{Human evaluation.}
Native-speaker participants for the paired comparison test and the naturalness MOS study were recruited through Prolific for English and Chinese and through Lancers for Japanese, and were paid above Prolific's minimum rate or, on Lancers, more than twice the Tokyo minimum wage (Appendix~\ref{app:he_recruitment}). Conditions were blinded, presentation order was randomized, and complete playback was required. The paired comparison test included catch trials, and no participant met the pre-specified exclusion criteria. We analyzed responses using participant identifiers solely for assignment, deduplication, and quality control, and collected no personally identifying information.

\paragraph{Artifact use and licensing.}
Apart from the weight-editing baselines of Appendix~\ref{app:pronunciation_baselines}, all third-party models, datasets, and evaluation tools used in this work are used without modification and remain subject to the licenses and terms of use of their original distribution sources. Apart from open-source code included under its own license, only the sample page described below and the ja-tts-g2p-bench items of Appendix~\ref{app:ja_reading_benchmark} redistribute third-party material, as the respective terms permit: one JSUT recording, the Japanese voice prompt, and Japanese sentences from ja-tts-g2p-bench and Wikipedia. Our released artifacts are intended for research and reproducibility purposes and do not supersede the original access conditions of any third-party resources. The released repository (\releaseurl) contains the generation, scoring, and analysis code, the container definitions and the requirements specification, the benchmark item lists, and the pre-computed evaluation outputs, together with the listening-test stimulus manifests, the item-level aggregates, the verbatim participant instructions in English, Japanese, and Chinese, and the interface screenshots. It also hosts a page of speech samples selected from the listening-test stimuli (\url{https://kuri-san-jp.github.io/speechsplice/}), released under the terms of LibriTTS, JSUT, and the Databaker corpus, from which the voice-cloning prompts derive, and of ja-tts-g2p-bench and Wikipedia, from which the Japanese same-surface items derive.

\paragraph{Data screening and documentation.}
We wrote the benchmark entity lists by hand from widely known public proper nouns, and we use them for research and evaluation purposes. Because proper nouns may include names of public individuals, we treated them as potentially identifying information and inspected the collected entity list to remove entries containing private personal information beyond public names, offensive expressions, or clearly inappropriate content. Human-evaluation records were kept separate from the benchmark entities and used only for the analysis of the paired comparison test and the MOS study and for quality control.

The automatic benchmark consists of three language splits---English, Japanese, and Chinese---with two sets of 300 entity-replacement pairs and one set of 300 two-entity sentences per language (Section~\ref{sec:experimental_setup}). The other-system evaluation uses the English 300-pair split of Table~\ref{tab:cross_lingual_overview}, the paired comparison test uses 50 name items per language plus 50 Japanese same-surface items, and the MOS study uses 50 pairs each in English, Chinese, and Japanese. The artifacts created in this work are intended for research and reproducibility purposes.

\paragraph{AI assistant disclosure.}
The authors consulted AI assistants for language polishing and checklist preparation. All AI-assisted text was reviewed and edited by the authors, who take full responsibility for the paper content. AI assistants did not generate experimental results or substitute for scientific analysis.

\bibliography{references}

\appendix
\setcounter{table}{0}
\renewcommand{\thetable}{A\arabic{table}}
\setcounter{figure}{0}
\renewcommand{\thefigure}{A\arabic{figure}}

\section{Acoustic Cross-Attention Mask Ablation}
\label{app:crossattn_acoustic}
\label{app:ja_token_subset}

Table~\ref{tab:crossattn_mask} gives the full four-condition ablation
summarized in Section~\ref{sec:convergence_results}.
Entity spans come from Whisper word-level timestamps \citep{radford2023robust},
subdivided to characters for languages without spaces. MEL
Similarity is the log-mel cosine similarity after time-averaging over the entity
span. MCD is variance-normalized and computed via dynamic time warping (DTW)
\citep{sakoe1978dynamic} over mel-frequency cepstral coefficient (MFCC) frames \citep{davis1980comparison} with $c_0$
dropped; the normalization raises the scale.
Restricting JA to pairs with matched token counts ($\Delta_{\mathrm{tok}}{=}0$) gives the same result: BL$_{\mathrm{mask}}$ and INJ$_{\mathrm{mask}}$ again yield MCD~$=0$ with nearly identical MEL Similarity.

\begin{table}[!t]
\centering
{\small
\setlength{\tabcolsep}{3pt}
\renewcommand{\arraystretch}{1.02}
\begin{tabular}{clcc}
\toprule
 & & \multicolumn{2}{c}{Cross-attention} \\
\cmidrule(lr){3-4}
Lang. & Metric & Intact & \makecell{Blocked\\at entity} \\
\midrule
EN & MCD (dB) $\uparrow$ & $\mathbf{35.31 \pm 0.27}$ & $0.00 \pm 0.00$ \\
   & \makecell{MEL\\Similarity $\downarrow$} & $\mathbf{0.88 \pm 0.01}$ & $1.00 \pm 0.00$ \\
JA & MCD (dB) $\uparrow$ & $\mathbf{31.73 \pm 0.77}$ & $0.00 \pm 0.00$ \\
   & \makecell{MEL\\Similarity $\downarrow$} & $\mathbf{0.94 \pm 0.01}$ & $1.00 \pm 0.00$ \\
ZH & MCD (dB) $\uparrow$ & $\mathbf{35.37 \pm 0.39}$ & $0.00 \pm 0.00$ \\
   & \makecell{MEL\\Similarity $\downarrow$} & $\mathbf{0.86 \pm 0.02}$ & $1.00 \pm 0.00$ \\
\bottomrule
\end{tabular}}
\caption{Acoustic cross-attention mask ablation on T5Gemma-TTS with length-preserving, mean-pooled injection (Section~\ref{sec:convergence_results}). The Intact condition compares BL--INJ, and the Blocked condition compares BL$_{\mathrm{mask}}$--INJ$_{\mathrm{mask}}$, the same pair synthesized with cross-attention to the entity positions blocked, on EN, JA, and ZH entity spans from Whisper word timestamps. Blocking cross-attention to the entity positions removes the BL--INJ difference in all three languages. Values are means with 95\% normal-approximation CI half-widths. These metrics capture acoustic difference, not target-direction accuracy.}
\label{tab:crossattn_mask}
\end{table}

\section{Phoneme-Level Verification of the Cross-Attention Mask Ablation (JA)}
\label{app:ja_crossattn_phoneme}

The acoustic ablation of Section~\ref{sec:convergence_results} reports MCD and MEL Similarity for the four cross-attention mask conditions in
Table~\ref{tab:crossattn_mask} (Appendix~\ref{app:crossattn_acoustic}). In the blocked-cross-attention condition, EN, JA, and ZH yield MCD~$=0$
with nearly identical MEL Similarity, indicating that the acoustic effect of injection is removed when
entity-position cross-attention is blocked. We complement this acoustic evidence with a phoneme-level
verification using 300 JA same-surface pairs, in which the written sentence is unchanged and only the intended reading differs; as a cross-check, we also use 300 EN same-surface pairs.

\paragraph{Procedure.}
For each pair, we generate five speech signals: $\mathrm{BL}$, $\mathrm{INJ}$, $\mathrm{BL}_{\mathrm{mask}}$,
$\mathrm{INJ}_{\mathrm{mask}}$, and a target-reading reference $\mathrm{TGT}$, obtained by directly
synthesizing the target-reading sentence for the analysis in Table~\ref{tab:ja_crossattn_phoneme}.

  \begin{table}[!t]
  \centering
  \footnotesize
  \setlength{\tabcolsep}{1.3pt}
  \begin{tabular}{@{}ccccc@{}}
  \toprule
   & \multicolumn{2}{c}{Intact} & \multicolumn{2}{c}{Blocked at entity} \\
  \cmidrule(lr){2-3}\cmidrule(lr){4-5}
  Lang. & BL & INJ & BL$_{\mathrm{mask}}$ & INJ$_{\mathrm{mask}}$ \\
  \midrule
  \multicolumn{5}{@{}l}{\textit{Target-match rate (entity PER\,$<$\,0.3), \%}~$\uparrow$} \\
  JA & \makecell{$20.00$\\$(+4.89/{-}4.13)$} & \makecell{$\mathbf{38.00}$\\$\mathbf{(+5.61/{-}5.31)}$} & \makecell{$0.00$\\${+}1.26$} & \makecell{$0.00$\\${+}1.26$} \\
  EN & \makecell{$1.00$\\$(+1.90/{-}0.66)$} & \makecell{$2.00$\\$(+2.29/{-}1.08)$} & \makecell{$0.00$\\${+}1.26$} & \makecell{$0.00$\\${+}1.26$} \\
  \midrule
  \multicolumn{5}{@{}l}{\textit{Mean entity PER vs.\ target-reading reference}~$\downarrow$} \\
  JA & \makecell{$0.569$\\$\pm 0.297$} & \makecell{$\mathbf{0.416}$\\$\mathbf{\pm 0.273}$} & \makecell{$0.885$\\$\pm 0.097$} & \makecell{$0.885$\\$\pm 0.103$} \\
  EN & \makecell{$0.898$\\$\pm 0.380$} & \makecell{$0.917$\\$\pm 0.416$} & \makecell{$1.222$\\$\pm 0.480$} & \makecell{$1.222$\\$\pm 0.480$} \\
  \bottomrule
  \end{tabular}
  \caption{Phoneme-level cross-attention mask ablation on T5Gemma-TTS. Unlike Table~\ref{tab:crossattn_mask}, injection uses the splice of Eq.~\eqref{eq:splice}, which is a replacement in place for all EN pairs and for the 73 JA pairs whose source and target spans have the same number of tokens. In JA, INJ moves the entity pronunciation toward the target reading with cross-attention intact but not with it blocked. Target-match rate denotes entity PER$<0.3$ against the target-reading reference. Values are percentages with Wilson 95\% CIs as $(+\mathrm{upper}/{-}\mathrm{lower})$ (top) and means $\pm$ standard deviations (bottom). For zero proportions, only the nonnegative upper bound of the Wilson interval is shown.}
  \label{tab:ja_crossattn_phoneme}
  \end{table}

\paragraph{Results.}
Table~\ref{tab:ja_crossattn_phoneme} follows a matched-baseline comparison logic: each intervention is
evaluated against its matched baseline. In JA, intact
cross-attention yields a modest target-directed improvement at the phoneme level, with INJ improving over BL
in both target-match rate and PER relative to the target reference. When entity-position cross-attention is
blocked, BL$_{\mathrm{mask}}$ and INJ$_{\mathrm{mask}}$ yield the same target-match rate and nearly the same
PER, indicating that the injected encoder states no longer affect the entity pronunciation. In EN,
INJ does not detectably improve over BL, and the two blocked conditions are identical.

\paragraph{Interpretation.}
Together with the acoustic MCD and MEL Similarity results in Table~\ref{tab:crossattn_mask},
Table~\ref{tab:ja_crossattn_phoneme} supports the transmission-path interpretation. The JA intact condition
shows that the transmitted signal can move the output toward the target at the phoneme level, whereas blocking
entity-position cross-attention removes both acoustic and phoneme-level differences. Thus,
Table~\ref{tab:ja_crossattn_phoneme} should be interpreted as evidence for the cross-attention pathway rather than
as a standalone pronunciation-control benchmark.

\section{Japanese 160-Item Reading Benchmark}
\label{app:ja_reading_benchmark}

\paragraph{Data and synthesis.}
Table~\ref{tab:ja160} summarizes the Japanese reading benchmark, which contains 160 adversarial contextual items in eight categories. Each item provides a kanji surface, the contextually intended reading, a competing reading, and full-sentence expected/alternative reading stimuli. We synthesized BL, a correct-reading reference, and a wrong-reading reference with frozen T5Gemma-TTS, JSUT voice cloning, seed 42, and the standard sampling configuration of top-$k$ sampling with $k{=}30$, top-$p$ sampling with $p{=}0.9$, and temperature $0.8$. For injection, the intended katakana form was encoded and its complete entity span was spliced into the source encoder output.

\paragraph{Reading decision.}
\texttt{Parakeet-Inc/\allowbreak furigana\_\allowbreak whisper\_\allowbreak small\_\allowbreak jsut}\urlfootnote{\texttt{Parakeet-Inc/\allowbreak furigana\_\allowbreak whisper\_\allowbreak small\_\allowbreak jsut}}{https://huggingface.co/Parakeet-Inc/furigana_whisper_small_jsut} directly transcribed speech to kana. After hiragana normalization, a transcript containing only the intended or only the competing reading was labeled accordingly. Remaining cases were assigned by lower full-sentence kana CER; differences below $0.02$ were labeled ambiguous. Five ambiguous BL items were excluded from the correction denominator.

\begin{table}[!t]
\centering
\footnotesize
\setlength{\tabcolsep}{2pt}
\begin{tabular}{@{}lrrrr@{}}
\toprule
Category & Items & \makecell{BL\\correct} & \makecell{BL\\wrong} & \makecell{Corrected\\/ wrong} \\
\midrule
Heteronym & 43 & 30 & 12 & 12/12 \\
Contextual numeral & 34 & 25 & 8 & 8/8 \\
Drama-specific reading & 24 & 17 & 7 & 7/7 \\
Okurigana & 35 & 23 & 12 & 12/12 \\
Jukujikun & 6 & 4 & 1 & 1/1 \\
Verb conjugation & 6 & 4 & 1 & 1/1 \\
Mixed on/kun compounds & 6 & 5 & 1 & 1/1 \\
Domain-dependent on/kun & 6 & 2 & 3 & 3/3 \\
\midrule
Total & 160 & 110 & 45 & 45/45 \\
\bottomrule
\end{tabular}
\caption{Japanese contextual-reading benchmark. BL correct and BL wrong count the items that baseline synthesis read with the intended and the competing reading, respectively, leaving out ambiguous readings; Corrected / wrong counts the BL-wrong items that splice injection read with the intended reading. Every strict BL misreading was corrected by splice injection, and every corrected output passed the direct containment criterion rather than the CER fallback.}
\label{tab:ja160}
\end{table}

The benchmark is intentionally enriched for low-frequency readings; BL wrong $=45/160$ should not be interpreted as a natural-text error rate. To validate the decision pipeline, we also scored 320 references with known polarity (160 intended-reading and 160 competing-reading syntheses). It returned the expected label for 317/320 (99.06\%), made no polarity reversal, and marked the remaining three cases ambiguous.

\section{Human Evaluation of the Same-Surface Correction Set}
\label{app:same_surface_human}

The same-surface set of the paired comparison test contains 50 Japanese items in which the baseline misreads the target word while the written sentence stays unchanged. Thirty-eight items come from \texttt{ja-tts-g2p-bench}\urlfootnote{\texttt{kawajiri-tellernovel/ja-tts-g2p-bench}}{https://huggingface.co/datasets/kawajiri-tellernovel/ja-tts-g2p-bench}, a Japanese kanji-reading benchmark for TTS released under CC BY-SA 4.0, which also supplies the 160 items of Appendix~\ref{app:ja_reading_benchmark}. The items were curated before data collection: starting from the strict BL errors of that benchmark, we replaced items whose question was not uniquely answerable, whose audio was degenerate, or whose word duplicated another item, and filled the set to 50 with difficult place names and surnames taken from Wikipedia\urlfootnote{Wikipedia}{https://www.wikipedia.org/} articles on hard-to-read Japanese place names and surnames, whose misreading and correction were verified automatically. Stimuli were synthesized with frozen T5Gemma-TTS and a JSUT\urlfootnote{JSUT corpus}{https://sites.google.com/site/shinnosuketakamichi/publication/jsut} voice-cloning prompt, following Appendix~\ref{app:ja_reading_benchmark}. The final composition is 9 heteronyms, 8 contextual numerals, 9 okurigana items, 6 drama-specific readings, 7 place names, 5 rare surnames, and 6 items from smaller categories.

The protocol was identical to the name sets in Section~\ref{sec:eval_metrics}: 20 native-speaker participants, two 25-item blocks, three catch trials each, and ten judgments per item. Table~\ref{tab:same_surface_results} reports the outcome. Participants chose the injected clip in almost every judgment, the item-level majority reached the target for all 50 items, and no participant met an exclusion criterion.

\begin{table}[!t]
\centering
\footnotesize
\begin{tabular}{@{}lr@{}}
\toprule
Measure & Value \\
\midrule
Participants & 20 \\
Items & 50 \\
Judgments & 500 \\
\midrule
INJ selected & 493 (98.6\%) \\
BL selected & 3 (0.6\%) \\
\emph{Neither} selected & 4 (0.8\%) \\
\midrule
INJ-selection rate (95\% CI) & 98.6 (97.4--99.6) \\
GEE $z$ ($p$) & 10.37 ($<10^{-16}$) \\
Item-level majority reaching target & 50/50 \\
Catch-trial pass rate & 100\% \\
Participants excluded & 0 \\
\bottomrule
\end{tabular}
\caption{Human evaluation of the Japanese same-surface correction set. Participants chose the injected clip in almost every judgment. The
INJ-selection rate is the unconditional rate with item-level bootstrap
confidence intervals, tested against chance with the same binomial GEE as
Table~\ref{tab:cross_lingual_overview}.}
\label{tab:same_surface_results}
\end{table}

\section{Conventional Pronunciation-Control Baselines}
\label{app:pronunciation_baselines}

Table~\ref{tab:pronunciation_baselines} compares SpeechSplice with interventions applied to text or model weights on the 300 Japanese pairs of the cross-architecture benchmark, using common entity spans, speaker references, and an ASR pipeline across conditions. Baseline synthesized the original sentence; G2P Baseline replaced the entity with an automatically predicted reading; Text Hint retained the entity and appended the target surname, written in kanji, in parentheses; and ROME/MEMIT edited encoder weights with Rank-One Model Editing (ROME) \citep{meng2022rome} and Mass-Editing Memory in a Transformer (MEMIT) \citep{meng2023memit}. SpeechSplice injected target entity states in every sample. All five conditions are synthesized with voice cloning and scored on ASR transcripts, so Target Reading Match (TRM, defined below) is not comparable with the PIB of Table~\ref{tab:cross_arch_main}, which compares entity phoneme error rates of speech synthesized without a voice prompt.

TRM marks a sample correct when the normalized target reading appears in at least one path of a multi-path pronunciation extractor applied to the ASR transcript. G2P-CER Match also accepts a sample when some window of the pykakasi reading of the ASR transcript, as long as the target reading, lies within a CER of $0.30$ of it:
\begin{align}
\mathrm{TRM} &= \frac{1}{N}\sum_{n=1}^{N}
  \mathbf{1}[\,\mathrm{norm}(r_n^{*}) \in P(\hat{y}_n)\,] \label{eq:trm}\\
\mathrm{G2P\text{-}CER} &= \frac{1}{N}\sum_{n=1}^{N} \mathbf{1}[\,t_n \lor{} \notag\\
&\quad \min_{w}\mathrm{CER}(g_{n,w},\, \rho_n) \leq 0.30\,] \label{eq:g2pcer}
\end{align}
Here $\hat{y}_n$ in Eqs.~\eqref{eq:trm} and~\eqref{eq:g2pcer} is the ASR transcript of sample $n$, $r_n^{*}$ is its target reading, $N$ is the number of samples, $t_n$ is the indicator inside Eq.~\eqref{eq:trm}, $P(\hat{y})$ is a multi-path pronunciation extractor that combines pykakasi readings and the raw transcript, $\rho_n = \mathrm{norm}(r_n^{*})$, and $g_{n,w}$ is a window of $\mathrm{norm}(\mathrm{G2P}(\hat{y}_n))$ as long as $\rho_n$, where $\mathrm{G2P}$ is the pykakasi reading. Values are percentages with Wilson 95\% intervals, shown as $(+\mathrm{upper}/{-}\mathrm{lower})$.

\begin{table}[!t]
\centering
\footnotesize
\setlength{\tabcolsep}{3pt}
\begin{tabular}{lcc}
\toprule
Condition & TRM (\%)$\uparrow$ & \makecell{G2P-CER\\Match (\%)$\uparrow$} \\
\midrule
Baseline & $0.00~{+}1.26$ & $0.00~{+}1.26$ \\
G2P Baseline & $0.00~{+}1.26$ & $0.00~{+}1.26$ \\
Text Hint & \makecell{$51.67$\\$(+5.60/{-}5.64)$} & \makecell{$52.33$\\$(+5.59/{-}5.65)$} \\
ROME/MEMIT & $0.00~{+}1.26$ & $0.00~{+}1.26$ \\
\midrule
SpeechSplice & \makecell{$\mathbf{98.67}$\\$\mathbf{(+0.81/{-}2.04)}$} & \makecell{$\mathbf{99.00}$\\$\mathbf{(+0.66/{-}1.90)}$} \\
\bottomrule
\end{tabular}
\caption{Comparison with text- and weight-level pronunciation-control baselines on the Japanese pairs of the cross-architecture benchmark. Direct encoder-output intervention recovers the target reading more often than text rewriting, text hints, or encoder weight editing on this split. For zero proportions, only the nonnegative upper bound of the Wilson interval is shown.}
\label{tab:pronunciation_baselines}
\end{table}

The null G2P result is specific to this construction: source and target are different entities, so an ordinary front end that preserves input content cannot produce the target entity. Text Hint partially exposes the target reading, whereas upstream weight editing does not recover the target reading in any sample.

\section{Full-Sentence CER}
\label{app:cer}

Whisper-INJ checks whether the target reading appears in the transcript, so it
reduces each sample to a binary outcome. As a continuous check we compute
the CER of the ASR transcripts against the source and target sentences; for EN
these are the Whisper medium.en transcripts of Appendix~\ref{app:compute}. Table~\ref{tab:cer}
reports the four combinations for the 300-pair benchmarks.

The baseline transcript sits close
to the source sentence and far from the target one, and injection reverses
both distances in all three languages. This holds under a measure that reads
the whole sentence rather than a single substring, so it does not inherit the
blind spots of containment matching. Injection also leaves the rest of the
sentence intact: the injected CER against the target is close to the baseline
CER against the source in JA and ZH. For English the same comparison shows a
larger gap.

Each language uses its own recognizer and normalization, so the CER values
are not comparable across languages. English lowercases and keeps
alphanumerics, Japanese keeps katakana after converting the reference with
pykakasi, and Chinese keeps Han characters. Confidence intervals are
$1.96\times$ the standard error over the 300 pairs.

\begin{table}[!t]
\centering
\footnotesize
\begin{tabular}{@{}lccc@{}}
\toprule
Transcript vs.\ reference & EN & JA & ZH \\
\midrule
BL vs.\ source $\downarrow$ & $0.008$ & $0.047$ & $0.041$ \\
BL vs.\ target $\uparrow$ & $0.250$ & $0.209$ & $0.245$ \\
INJ vs.\ target $\downarrow$ & $0.016$ & $0.050$ & $0.045$ \\
INJ vs.\ source $\uparrow$ & $0.255$ & $0.215$ & $0.246$ \\
\bottomrule
\end{tabular}
\caption{Full-sentence CER on the name-replacement pairs of Table~\ref{tab:cross_lingual_overview}
(95\% CI half-widths are $0.005$ to $0.018$). The baseline transcript is close to the source sentence and the injected transcript to the target sentence in every language. Lower is better when the
transcript should match the reference, and higher is better for the two
control rows. Recognizers and normalization differ by language, so values are
not comparable across columns.}
\label{tab:cer}
\end{table}

\section{Human Evaluation Details}
\label{app:human_eval_details}

This appendix documents the two listening studies of
Section~\ref{sec:eval_metrics}: recruitment and payment, consent and
data handling, the full instructions shown to
participants, and the interface and quality controls.

\subsection{Statistical test of the paired comparison test}
\label{app:he_stats}
INJ-selection rates were tested against chance (50\%) using a binomial generalized estimating equation (GEE) \citep{liang1986longitudinal}, with items treated as clusters, an exchangeable working correlation, and cluster-robust standard errors (two-sided, Bonferroni-corrected $\alpha=0.0125$). For ZH every judgment selected INJ, so the GEE is completely separated and its Wald $z$ is not defined; we report the pre-specified fallback, an exact binomial test on the 50 item-level majorities.

\subsection{Recruitment and payment}
\label{app:he_recruitment}

English and Chinese participants were recruited on Prolific\urlfootnote{Prolific}{https://www.prolific.com/} and Japanese
participants on Lancers\urlfootnote{Lancers}{https://www.lancers.jp/}, a Japanese crowdsourcing platform. The Prolific
filters required English (paired comparison test and MOS study in EN) or Mandarin
(ZH) as the first language and an approval rate of 95--100\%; the English
studies were limited to residents of the United Kingdom and the United
States, and the Chinese studies had no residence restriction. The
Japanese recruitment text required native speakers of Japanese. The
recruitment text disclosed the task content, the expected duration, and
the payment before enrollment, and asked participants to use
headphones. Platform identifiers and per-study exclusion lists
prevented duplicate participation within a study.

All realized hourly rates were above Prolific's 8.00~USD-per-hour minimum, and
the Lancers rates were more than twice the Tokyo minimum wage
(1,226~JPY per hour). Every analyzed submission was
approved and paid. Ten English paired-comparison-test submissions received the
wrong stimulus block through a study-configuration error on our side;
they were paid in full and excluded from analysis.

\subsection{Consent and data handling}
\label{app:he_consent}

Participation followed the standard consent model of both platforms:
the recruitment text disclosed the task, its duration, the payment, and
the research purpose, and only workers who accepted these terms took
part. Participation was voluntary and could be abandoned at any time.
We did not use a separate consent form. At login we collected the platform participant identifier, an age band,
a gender category, and the listening device. On Prolific the identifier
was filled in automatically from the study URL and no name field was
shown, following the platform's prohibition on collecting real names.
The Lancers form carried one further field labelled as a name.
Requesters on Lancers see only an anonymized worker identifier, so
participants entered that identifier in both fields, and both served
solely for completion-code and payment matching. We stored no real
names.
We collected no contact information, no audio of participants, and no
free-text responses. Responses are stored on a research server
accessible only to the researchers, are reported only in aggregate, and
participant-level raw responses are not released. The identifiers
served solely for assignment, deduplication, quality control, and
payment matching.

\subsection{Ethics review}
\label{app:he_ethics}

The authors' organization does not have an institutional review board,
and no formal approval or exemption determination was obtained. The
studies are minimal-risk online listening tasks: all stimuli are
synthetic speech generated by the authors, the tasks involve no
deception and no sensitive content, both platforms admit adults only,
participation was voluntary with the task disclosed upfront, no
personally identifying information was collected, and payment exceeded
Prolific's minimum rate or, on Lancers, the Tokyo minimum wage.

\subsection{Instructions shown to participants}
\label{app:he_instructions}

The English instructions below are verbatim. The Japanese and Chinese
instructions of both studies are reproduced verbatim, with English
glosses, in \texttt{human\_eval/\allowbreak instructions\_\allowbreak verbatim.md}
of the released repository (\releaseurl). The English and Chinese
instructions of the paired comparison test and the Japanese and Chinese MOS
instructions are the authors' translations and were not independently
validated by professional translators.

\paragraph{Paired comparison test (name sets).}
The English version, titled ``Pronunciation Matching Test'':
\begin{enumerate}\itemsep0pt
\item You will hear two audio clips in sequence.
\item The first voice $\rightarrow$ 2 seconds of silence $\rightarrow$ the second voice plays automatically.
\item A name is shown at the top of the screen. Choose which clip says that name.
\item Judge by how the name is pronounced. Do not consider the naturalness or audio quality of the voice.
\item If neither clip says the displayed name, choose ``Neither says the displayed name''.
\item You will start with 2 practice questions. Practice answers are not included in the results.
\end{enumerate}
Each trial asked ``Which one says the displayed name?'' with the
choices \emph{First voice} / \emph{Neither says the displayed name} /
\emph{Second voice}. The practice screens added ``This is practice.
Your answers are not included in the results.''

The Japanese and Chinese versions (both titled ``Name Listening Test''
in our gloss) present the same six items, question, choices, and
practice note, except that Japanese item~4 reads ``Judge only by the
sequence of kana (the reading). Do not consider pitch accent or the
naturalness or audio quality of the voice.''

\paragraph{Paired comparison test (Japanese same-surface set).}
The same-surface study used the Japanese instructions above with four
changes (glossed): the title was ``Word Reading Listening Test'', item~3
read ``For a word in the sentence, choose the clip that pronounces it
with the reading shown on the screen'', item~5 pointed to the center
button, and each trial asked ``Which one pronounces it with the displayed
reading?''

\paragraph{Naturalness MOS.}
The English instructions: ``You will evaluate the naturalness and
acoustic quality of speech samples. First, please complete this short
practice session.'' followed by the steps ``Click the `Play' button to
listen to the audio. / Listen to the full audio clip. / Rate the
naturalness and acoustic quality on a 5-point scale. / Click `Next' to
proceed to the next sample.'' A highlighted warning read: ``Note: Do
NOT judge pronunciation correctness. Please do not evaluate whether
proper nouns are pronounced correctly. Rate only the voice quality,
fluency, and naturalness of the speech.'' The scale was defined as 5 =
Completely natural, high-quality speech; 4 = Slightly unnatural, but
high quality; 3 = Somewhat unnatural, but understandable; 2 =
Noticeably unnatural, difficult to listen to; 1 = Very unnatural, low
quality, with the button labels 1~Bad / 2~Poor / 3~Fair / 4~Good /
5~Excellent. Each trial repeated the warning and required listening to
the full clip before proceeding.

The Japanese and Chinese MOS instructions are the authors' translations
of the same text, including the warning and the five scale definitions.

\subsection{Interface and quality control}
\label{app:he_interface}

Both studies were conducted in a browser application with the flow: login with the
four fields of Appendix~\ref{app:he_consent}, instruction screen, two practice trials, main
trials, and a completion-code screen. In the paired comparison test the two
clips played automatically in sequence with two seconds of silence
between them, and the answer buttons stayed disabled until both clips
had played to the end; in the MOS study the next button stayed disabled
until the clip had played to the end. Item order was shuffled
server-side for every session, and the BL/INJ position in each
paired-comparison-test trial was assigned by a server-side uniform draw and logged
with the response. The interface showed no condition labels; the audio file names
carried the condition but were visible only by inspecting the page.

Catch trials reused genuine stimulus speech and displayed a third word
that neither clip speaks; the correct answer is the \emph{neither}
button. Pass rates were 100\% (EN), 98.33\% (JA names), 100\% (ZH), and
100\% (JA same-surface), and no participant failed more than one of the
three catch trials. The MOS study contained no catch trials.
In the MOS study the 100 clips per language were split into two
50-clip blocks of 25 pairs each, and ten participants rated each block in JA and ZH, and nine and eleven in EN,
which yields 1{,}000 ratings per language.

\section{Failure Cases Exposed by the Listening Study}
\label{app:failures}

Listener majorities and the Whisper judgments disagreed on eight
name items. Seven are recognition errors: the listeners heard the target reading
and the transcript did not contain it, because the recognizer wrote a
similar-sounding word instead. These items count as failures in the
automatic scores while the speech itself is correct, which is why we treat
the automatic scores as conservative.

The eighth item is the reverse case and the more informative one. The
injected clip realized the target reading but carried an extra mora, and
most participants chose \emph{neither} rather than accept it, while substring
matching found the target reading inside the longer form and scored the item
as a success. This is the failure mode that automatic scoring misses, and it
is why the paired comparison test offers the \emph{neither} option.

\section{Compute and Model Details}
\label{app:compute}

  \paragraph{Models and parameter counts.}
  All TTS and auxiliary models are used as publicly released frozen checkpoints, except that the ROME/MEMIT baselines of Appendix~\ref{app:pronunciation_baselines} edit encoder weights;
  no fine-tuning is performed. Table~\ref{tab:models} lists them.
  T5Gemma-TTS uses XCodec2 \citep{ye2025llasa} as its audio tokenizer and
  vocoder. We compute all PER and PIB values from the International Phonetic Alphabet (IPA) transcripts of
  \texttt{wav2vec2-xlsr-53-espeak-cv-ft}\urlfootnote{facebook/wav2vec2-xlsr-53-espeak-cv-ft}{https://huggingface.co/facebook/wav2vec2-xlsr-53-espeak-cv-ft} \citep{xu2022phoneme}.

\paragraph{Checkpoints and tools.}
The TTS checkpoints are
T5Gemma-TTS\urlfootnote{Aratako/T5Gemma-TTS-2b-2b}{https://huggingface.co/Aratako/T5Gemma-TTS-2b-2b},
SpeechT5\urlfootnote{microsoft/speecht5\_tts}{https://huggingface.co/microsoft/speecht5_tts},
LongCat-AudioDiT\urlfootnote{meituan-longcat/LongCat-AudioDiT-1B}{https://huggingface.co/meituan-longcat/LongCat-AudioDiT-1B},
Irodori-TTS\urlfootnote{Aratako/Irodori-TTS-v4.1-Small}{https://huggingface.co/Aratako/Irodori-TTS-v4.1-Small},
Kiseki-TTS\urlfootnote{telecomadm1145/Kiseki-TTS}{https://huggingface.co/telecomadm1145/Kiseki-TTS},
Qwen3-TTS\urlfootnote{Qwen/Qwen3-TTS-12Hz-1.7B-Base}{https://huggingface.co/Qwen/Qwen3-TTS-12Hz-1.7B-Base},
XTTS-v2\urlfootnote{coqui/XTTS-v2}{https://huggingface.co/coqui/XTTS-v2},
Llasa-3B\urlfootnote{HKUSTAudio/Llasa-3B}{https://huggingface.co/HKUSTAudio/Llasa-3B},
CSM-1B\urlfootnote{sesame/csm-1b}{https://huggingface.co/sesame/csm-1b}, and
Bark\urlfootnote{suno/bark}{https://huggingface.co/suno/bark}.
Speaker similarity uses
WavLM-SV\urlfootnote{microsoft/wavlm-base-plus-sv}{https://huggingface.co/microsoft/wavlm-base-plus-sv}.
Whisper large-v3 runs with greedy language-forced decoding. When no transcript of a voice prompt is given, T5Gemma-TTS voice cloning transcribes the prompt with Whisper large-v3-turbo\urlfootnote{openai/whisper-large-v3-turbo}{https://huggingface.co/openai/whisper-large-v3-turbo}. The JA
Whisper-INJ results in Table~\ref{tab:cross_lingual_overview} and in the
two-entity columns of Table~\ref{tab:cross_arch_main} use
Furigana Whisper\urlfootnote{Parakeet-Inc/furigana\_whisper\_small\_jsut}{https://huggingface.co/Parakeet-Inc/furigana_whisper_small_jsut},
and the English CER of Appendix~\ref{app:cer} uses a separate set of
transcripts from
Whisper medium.en\urlfootnote{openai/whisper-medium.en}{https://huggingface.co/openai/whisper-medium.en}.
We extract $F_0$ with
pyworld\urlfootnote{pyworld package}{https://github.com/JeremyCCHsu/Python-Wrapper-for-World-Vocoder}
Harvest \citep{morise2017harvest} over voiced frames. For language-specific text processing, we
applied task-appropriate phonemization and normalization tools, including
g2p\_en\urlfootnote{g2p\_en package (Kyubyong/g2p)}{https://github.com/Kyubyong/g2p},
pykakasi\urlfootnote{pykakasi package (miurahr/pykakasi)}{https://codeberg.org/miurahr/pykakasi}, and
pypinyin\urlfootnote{pypinyin package (mozillazg/python-pinyin)}{https://github.com/mozillazg/python-pinyin}.

  \begin{table}[!t]
  \centering
  \footnotesize
  \setlength{\tabcolsep}{3pt}
  \begin{tabularx}{\columnwidth}{@{}>{\raggedright\arraybackslash}p{0.38\columnwidth}>{\raggedright\arraybackslash}X>{\raggedleft\arraybackslash}p{0.14\columnwidth}@{}}
  \toprule
  Model & Role & \#Params \\
  \midrule
  T5Gemma-TTS-2b-2b & Primary TTS (encoder--decoder) & 5.31B \\
  SpeechT5 (TTS) & Comparison TTS (encoder--decoder) & 144M \\
  LongCat-AudioDiT-1B & Comparison TTS (encoder--decoder) & 1.26B \\
  Irodori-TTS-v4.1-Small & Comparison TTS (encoder--decoder) & 766M \\
  Kiseki-TTS & Comparison TTS (encoder--decoder) & 411M \\
  Qwen3-TTS-12Hz-1.7B & Comparison TTS (decoder-only) & 1.93B \\
  XTTS-v2 & Comparison TTS (decoder-only) & 441M \\
  Llasa-3B \citep{ye2025llasa} & Comparison TTS (decoder-only) & 3.41B \\
  CSM-1B & Comparison TTS (decoder-only) & 1.55B \\
  Bark & Comparison TTS (decoder-only) & 1.09B \\
  XCodec2 \citep{ye2025llasa} & Audio tokenizer / vocoder & 830M \\
  VoiceCraft & Zero-shot TTS & 840M \\
  EdiTTS & Score-based content replacement & 14.8M \\
  Whisper large-v3 & Multilingual ASR evaluation & 1.54B \\
  Whisper large-v3-turbo & Voice-prompt transcription & 809M \\
  Whisper medium.en & English CER transcripts & 762M \\
  wav2vec2-xlsr-53-espeak-cv-ft \citep{xu2022phoneme} & IPA phoneme transcription & 316M \\
  UTMOS & MOS predictor & 103M \\
  \bottomrule
  \end{tabularx}
  \caption{Models used in this work, their role, and their parameter count, computed
  from the released weights. For a TTS system the count excludes the audio codec,
  autoencoder, or vocoder that produces the waveform.}
  \label{tab:models}
  \end{table}

  \paragraph{Replaced tensor per system.}
  Table~\ref{tab:cross_arch_main} replaces one tensor per system, chosen as
  the text-side state the generator reads. In the five encoder--decoder systems this is the encoder output that
  reaches cross-attention: the encoder \texttt{last\_hidden\_state} in
  T5Gemma-TTS and in SpeechT5; in LongCat-AudioDiT the cross-attention input,
  which adds the layer-normalized embedding back to the layer-normalized final
  encoder state; in Irodori-TTS the text state after the $768{\rightarrow}512$
  projection and its normalization; and in Kiseki-TTS the encoder output after
  the final normalization, which is distributed to the cross-attention keys and
  values of the decoder. The last three are therefore not layer outputs
  themselves. In Irodori-TTS, replacing the last layer's output, as the layer
  sweep does, gives the same audio, since the normalizations and the
  projection act on each token separately. In the five decoder-only systems the replaced tensor is a
  hidden state on the self-attention sequence during prefill, taken at the
  entity-word rows: the output of GPT block 29 of 30 in XTTS-v2, of decoder
  layer 27 of 28 in Llasa-3B, of backbone layer 15 of 16 in CSM-1B, of semantic
  transformer layer 23 of 24 in Bark, and of talker layer 27 of 28 in
  Qwen3-TTS, counted as in Section~\ref{sec:eval_metrics}. Section~\ref{sec:layer_results} varies this choice.

  \paragraph{Generation length.}
  T5Gemma-TTS synthesizes a clip of a length fixed before decoding: its duration estimator sets the length in seconds from the input text and the voice prompt. For INJ we pass the length estimated from the source sentence, which is the length BL uses, so the two clips have the same duration without BL being synthesized first. The T5Gemma-TTS outputs of Tables~\ref{tab:cross_lingual_overview} and~\ref{tab:system_comparison} were generated this way.

  \paragraph{Compute.}
  All experiments are inference only and were conducted on a single NVIDIA A100 40GB GPU (the single-entity Irodori-TTS experiments on an H100 80GB). Synthesis took about 275 GPU-hours and scoring about 40 GPU-hours.

\section{Paired Differences and Speech Quality in the Two-Entity Test}
\label{app:multientity_quality}
\makeatletter\setlength{\@fptop}{0pt}\makeatother
\begin{table}[!t]
\centering
\footnotesize
\setlength{\tabcolsep}{1.2pt}
\begin{tabular}{lccccc}
\toprule
& & \multicolumn{3}{c}{UTMOS~$\uparrow$} & \\
\cmidrule(lr){3-5}
Model & Lang. & BL & \makecell{INJ\\(one)} & \makecell{INJ\\(both)} & \makecell{INJ (both)\\$-$ BL} \\
\midrule
\multicolumn{6}{l}{\emph{Encoder--decoder (cross-attention)}} \\
\multirow{3}{*}{T5Gemma-TTS} & EN & 3.92 & 3.93 & 3.90 & $-0.0145 \pm 0.0572$ \\
 & JA & 3.35 & 3.32 & 3.31 & $-0.0408 \pm 0.0475$ \\
 & ZH & 3.05 & 3.07 & 3.08 & $+0.0319 \pm 0.0960$ \\
SpeechT5 & EN & 4.20 & 4.19 & 4.15 & $-0.0574 \pm 0.0269$ \\
\multirow{2}{*}{\makecell[l]{LongCat-\\AudioDiT}} & EN & 3.45 & 3.46 & 3.47 & $+0.0147 \pm 0.0545$ \\
 & ZH & 2.55 & 2.53 & 2.52 & $-0.0337 \pm 0.0459$ \\
Irodori-TTS & JA & 3.77 & 3.76 & 3.77 & $-0.0037 \pm 0.0324$ \\
Kiseki-TTS & JA & 2.42 & 2.42 & 2.42 & $+0.0021 \pm 0.0496$ \\
\midrule
\multicolumn{6}{l}{\emph{Decoder-only (self-attention)}} \\
\multirow{3}{*}{Qwen3-TTS} & EN & 4.52 & 4.51 & 4.51 & $-0.0030 \pm 0.0026$ \\
 & JA & 4.36 & 4.36 & 4.36 & $-0.0003 \pm 0.0107$ \\
 & ZH & 3.73 & 3.72 & 3.72 & $-0.0101 \pm 0.0218$ \\
XTTS-v2 & EN & 4.23 & 4.23 & 4.23 & $-0.0007 \pm 0.0088$ \\
\multirow{2}{*}{Llasa-3B} & EN & 3.59 & 3.59 & 3.61 & $+0.0222 \pm 0.0364$ \\
 & ZH & 2.80 & 2.83 & 2.84 & $+0.0403 \pm 0.0415$ \\
CSM-1B & EN & 4.21 & 4.21 & 4.20 & $-0.0054 \pm 0.0253$ \\
\multirow{2}{*}{Bark} & EN & 3.67 & 3.67 & 3.67 & $+0.0019 \pm 0.0092$ \\
 & JA & 2.47 & 2.47 & 2.47 & $+0.0012 \pm 0.0123$ \\
\bottomrule
\end{tabular}
\caption{UTMOS of the two-entity sentences of Table~\ref{tab:cross_arch_main} for the baseline (BL) and for one or both entities replaced (INJ), on each model and language. The last column is the paired difference between INJ (both) and BL; $\pm$: 95\% CI. Replacing both entities leaves UTMOS close to the baseline in every condition, and the decreases in SpeechT5 EN and Qwen3-TTS EN are small.}
\label{tab:multientity_utmos}
\end{table}

\paragraph{Paired accuracy difference.}
To compare INJ (one) and INJ (both) in Table~\ref{tab:cross_arch_main} without treating the two slots of a sentence as independent, we average the two per-slot verdicts of each sentence and take the difference INJ (both) $-$ INJ (one) per sentence. The mean difference, in percentage points with a 95\% percentile bootstrap interval over the 300 sentences (10{,}000 resamples), is $-0.50$ $[-2.17, +1.17]$, $+0.33$ $[-2.33, +3.00]$, and $-0.83$ $[-4.67, +3.00]$ for T5Gemma-TTS EN, JA, and ZH; $+1.33$ $[-0.83, +3.50]$ for SpeechT5 EN; $-1.17$ $[-2.67, +0.33]$ and $-0.17$ $[-0.83, +0.50]$ for LongCat-AudioDiT EN and ZH; $+0.33$ $[-1.33, +1.83]$ for Irodori-TTS JA; and $-0.83$ $[-2.17, +0.50]$ for Kiseki-TTS JA. Every interval includes zero; we did not test for equivalence. In the decoder-only conditions both columns are zero for every sentence.

\paragraph{Speech quality.}
The two-entity columns of Table~\ref{tab:cross_arch_main} score pronunciation only. To check whether replacing two entities in one sentence lowers speech quality, we scored every clip of the two-entity test with the UTMOS predictor used for Table~\ref{tab:cross_lingual_overview}. INJ (one) averages the two clips of each sentence in which one entity is replaced. The last column is the mean paired difference between INJ (both) and BL over the 300 sentences of each condition. Table~\ref{tab:multientity_utmos} shows that this difference stays small in every condition; its interval excludes zero only for SpeechT5 EN and Qwen3-TTS EN, where UTMOS drops slightly. UTMOS is an automatic estimate, and we conducted no listening test on these clips.

\end{document}